\documentstyle[12pt,twoside]{article}

\def\greaterthansquiggle{\raise.3ex\hbox{$>$\kern-.75em\lower1ex\hbox{$\sim$}}}
\def\lessthansquiggle{\raise.3ex\hbox{$<$\kern-.75em\lower1ex\hbox{$\sim$}}}
\newcommand{\beq}{\begin{equation}}
\newcommand{\eeq}{\end{equation}}
\newcommand{\beqa}{\begin{eqnarray}}
\newcommand{\eeqa}{\end{eqnarray}}
\newcommand{\beqan}{\begin{eqnarray*}}
\newcommand{\eeqan}{\end{eqnarray*}}
\newcommand{\ba}{\begin{array}}
\newcommand{\ea}{\end{array}}

\def\nz{\ifmmode {I\hskip -3pt N} \else {\hbox {$I\hskip -3pt N$}}\fi}
\def\zz{\ifmmode {Z\hskip -4.8pt Z} \else
       {\hbox {$Z\hskip -4.8pt Z$}}\fi}
\def\qz{\ifmmode {Q\hskip -5.0pt\vrule height6.0pt depth 0pt
       \hskip 6pt} \else {\hbox
       {$Q\hskip -5.0pt\vrule height6.0pt depth 0pt\hskip 6pt$}}\fi}
\def\rz{\ifmmode {I\hskip -3pt R} \else {\hbox {$I\hskip -3pt R$}}\fi}
\def\cz{\ifmmode {C\hskip -4.8pt\vrule height5.8pt\hskip 6.3pt} \else
       {\hbox {$C\hskip -4.8pt\vrule height5.8pt\hskip 6.3pt$}}\fi}

\newtheorem{theorem}{Theorem}

\newtheorem{lemma}{Lemma}

\normalsize
\begin{document}

\begin{titlepage}
\begin{center}
{\Large \bf 
Slow decay of dynamical correlation functions for nonequilibrium 
quantum states
}\\[24pt]
Takayuki Miyadera\\
Department of Information Sciences \\
Tokyo University of Science \\
Noda City, Chiba 278-8510,
Japan
\vfill
{\bf Abstract} \\
\end{center}
A property of 
dynamical correlation functions for nonequilibrium states is discussed. 
We consider arbitrary dimensional quantum spin systems with 
local interaction and translationally invariant states with 
nonvanishing current over them. 
A correlation function 
between local charge and local Hamiltonian at different 
spacetime points 
is shown to exhibit slow decay. 
\vfill
\small
e-mail: miyadera@is.noda.tus.ac.jp
\end{titlepage}
\section{Introduction}
In this paper we treat nonequilibrium states of quantum spin systems.
When we speak of nonequilibrium states, there are 
two kinds of states, that is, states near equilibrium and 
far from equilibrium. 
As for the investigation of the former nonequilibrium states,
the main purpose is to understand how the states 
approach to an equilibrium state.
The latter are states which, on the other hand, can not be treated as 
perturbed equilibrium states and do not converge to any 
equilibrium states, whose properties are less known.
We, in this paper, study a property of the latter nonequilibrium states.
There have been works on the nonequilibrium states from various
points of view \cite{DKS,DLS,HoA,Ta,Shmz,Oji,Fr,JP,SST,M}. 
Among the various results, 
 it is known that nonequilibrium steady states exhibit slow
decay of equal-time correlation function
from approximate theories like fluctuating hydrodynamics
\cite{DKS}.
Recently an approach to the problem from exactly solvable models
\cite{DLS}
has been investigated and the 
expected behavior of equal-time correlation function was confirmed.
\par
In this paper, another type of slow decay of 
dynamical correlation function will be discussed. 
We consider locally interacting quantum 
spin systems and translationally invariant states with 
nonvanishing current over them. 
We do not impose any other conditions like stationarity of the 
states and do not ask how the nonequilibrium states 
are realized.
Since the states have
finite current at infinity, they cannot be locally
perturbed equilibrium states and thus are far from 
equilibrium.
We consider a correlation function 
between local charge and local Hamiltonian at different 
spacetime points. 
It will be shown in mathematically 
rigorous way that the correlation functions of the states
exhibit slow decay with respect to space and time. 
The way to show the claim is a generalization of the 
method employed in \cite{M}.

 The paper is organized as follows.
In section \ref{sec2}, we introduce our nonequilibrium states 
on quantum spin systems. In section \ref{sec3}, the main theorem
is proved.
\section{States with Nonvanishing Current on Quantum Spin Systems}
\label{sec2}
Let us consider a $d$-dimensional ($d=1,2,3,\cdots$) infinite quantum 
spin system (see e.g., \cite{BR}).
For simplicity, we assume nearest neighbor translationally 
invariant interaction. 
That is, for each pair of neighboring sites ${\bf x}$ and ${\bf y}$ 
(i.e.,${\bf x},{\bf y} \in {\bf Z}^d$ s.t. 
$|{\bf x}-{\bf y}|=1$), a self adjoint operator 
$h({\bf x},{\bf y})$ is defined and 
satisfies 
$\tau_{\bf z}\left( h({\bf x},{\bf y})\right)=
 h({\bf x}+{\bf z},{\bf y}+{\bf z})$  
for all ${\bf z} \in {\bf Z}^d$
where $\tau_{\bf z}$
stands for the translation operator.
Denoting fundamental vector 
with direction $i$ as ${\bf e}_i$, local Hamiltonian 
with respect to a finite region $\Lambda \subset {\bf Z}$
is defined as
\begin{eqnarray*}
H_{\Lambda}=\sum_{r=1}^d \sum_{\{{\bf x},{\bf x}+{\bf e}_r\}
\subset \Lambda} h({\bf x},{\bf x}+{\bf e}_r).
\end{eqnarray*} 
In this paper we employ the Heisenberg picture and 
the Hamiltonian defines time evolution of local operators, say $A$, 
by 
\begin{eqnarray*}
\alpha_t(A):=\lim_{\Lambda \to {\bf Z}^d} 
e^{iH_{\Lambda}t} A e^{-iH_{\Lambda}t},
\end{eqnarray*}
where the limit is taken with respect to 
norm topology (we put $\hbar =1$).
To discuss (electric) current, 
a charge $n({\bf x})$, a self-adjoint
operator, is assumed to be defined on each site ${\bf x}$ 
\cite{energycurrent}, 
and the charge on a finite region $\Lambda$ is denoted as
$N_{\Lambda}:=\sum_{{\bf x}\in {\Lambda}}n({\bf x})$.
It induces a symmetry transformation on each finite 
sublattice $\Lambda$, that is,
$[H_{\Lambda},N_{\Lambda}]=0$ holds.    
The current operator between sites ${\bf x}$ and ${\bf x}+{\bf e}_r$ 
is defined by $j({\bf x},{\bf x}+{\bf e}_r):=
i[n({\bf x}),h({\bf x},{\bf x}+{\bf e}_r)]
=-i[n({\bf x}+{\bf e}_r),h({\bf x},{\bf x}+{\bf e}_r)]$.
\\
The above seemingly abstract setting has physically interesting
examples. For instance, interacting fermion system is on the list. 
For each ${\bf x}\in {\bf Z}^d$, 
charge is defined as 
$n({\bf x}):=c_{{\bf x}}^{\ast }c_{{\bf x}}$ where 
$c_{\bf x}$ is a fermionic annihilation operator 
at site ${\bf x}$ and 
$h({\bf x},{\bf x}+{\bf e}_r)=-T(c_{{\bf x}+{\bf e}_r}^{\ast }
c_{{\bf x}}+c_{{\bf x}}^{\ast }c_{{\bf x}+{\bf e}_r})
+V n({\bf x})n({\bf x}+{\bf e}_r)$ 
gives a nearest neighbor Hamiltonian where 
$T$ and $V$ are real constants. 
The current operator is calculated as 
$j({\bf x},{\bf x}+{\bf e}_r)=
iT(c_{{\bf x}+{\bf e}_r}^{\ast}c_{{\bf x}}-c_{{\bf x}}^{\ast }
c_{{\bf x}+{\bf e}_r})$. 
The Heisenberg model can be another example. 
On each site ${\bf x} \in {\bf Z}$, 
spin operators $\left( S_{{\bf x}}^{(1)},
S_{{\bf x}}^{(2)},S_{{\bf x}}^{(3)} \right)$ 
satisfying usual 
commutation relations for angular momentum 
live on.
$h({\bf x},{\bf x}+{\bf e}_r)
:=S_{{\bf x}}^{(1)}S_{{\bf x}+{\bf e}_r}^{(1)}
+S_{{\bf x}}^{(2)}S_{{\bf x}+{\bf e}_r}^{(2)}+\lambda
S_{{\bf x}}^{(3)}S_{{\bf x}+{\bf e}_r}^{(3)}$ ($\lambda$ is 
a real constant) 
and $n({\bf x}):=S_{{\bf x}}^{(3)}$ leads the current $%
j({\bf x},{\bf x}+{\bf e}_r)
=-S_{{\bf x}}^{(2)}S_{{\bf x}+{\bf e}_r}^{(1)}
+S_{{\bf x}}^{(1)}S_{{\bf x}+{\bf e}_r}^{(2)}$. 
\\
In this paper, we consider translationally invariant 
states with nonvanishing current at time $t=0$. 
Without loss of generality, we can fix the first axis ${\bf e}_1$ as 
a direction of nonvanishing current.
That is, we consider translationally invariant 
states which satisfy 
\begin{eqnarray*}
\langle j({\bf 0},{\bf e}_1)\rangle \neq 0.
\end{eqnarray*}
Since our conditions are weak, it surely contains 
a several physically interesting models \cite{HoA,Ta}.
However, we do not impose any other conditions like
stationarity and stability,
and therefore some states might be hardly realized in 
real physical situation \cite{real}. 
The crucial point is that the states have nonvanishing 
current at infinity and they are not, say, locally perturbed 
equilibrium states.
\section{Slow decay of correlation function}\label{sec3}
In derivation of slow decay of correlation functions for 
continuous symmetry breaking equilibrium states 
(Goldstone theorem),
it is crucial to represent an order parameter 
by a commutator \cite{Req,LFW}. 
Our discussion can be considered as a
nonequilibrium analogue of \cite{Req}.
For nonequilibrium states,
the following is an important observation. For 
finite sublattices $V$ and $\Lambda$ satisfying 
$V\supset \Lambda$, 
\begin{eqnarray}
[
-i\sum_{{\bf x} \in V}x_1n({\bf x}),H_{\Lambda}]
=\sum_{\{{\bf y},{\bf y}+{\bf e}_1\}
\subset \Lambda}j({\bf y},{\bf y}+{\bf e}_1)
\label{1st}
\end{eqnarray}
holds where $x_1$ is a component of ${\bf x}=(x_1,x_2,\cdots,x_d)
\in {\bf Z}^d$.  
Hereafter, all the finite sublattices are assumed to 
be $d$-dimensional cubes, and we use simplified notations 
$H_R:=H_{[-R,R]^d}$ for positive integer $R$ where 
$[-R,R]^d:=\{{\bf x}=(x_1,x_2,\cdots,x_d)\in {\bf Z}^d|
-R \leq x_j \leq R \ \mbox{for all}\ j=1,2,\cdots,d\}$.
Taking expectation value of (\ref{1st}) with respect to 
a translationally invariant state and we put 
$\Lambda =[-R,R]^d$, take $W>R$ and divide (\ref{1st}) 
by $|\{{\bf y}|\{{\bf y},{\bf y}+{\bf e}_1\} \subset [-R,R]^d \}|
=2R(2R+1)^{d-1}$. Then we obtain 
\begin{eqnarray*}
\frac{1}{2R(2R+1)^{d-1}}
\langle [
-i\sum_{{\bf x} \in [-W,W]^d}x_1n({\bf x}),H_{R}]
\rangle 
=\langle j({\bf 0},{\bf e}_1)\rangle,
\end{eqnarray*}
where we have used the translational invariance 
of the state.
By letting the size of cubic lattices in 
the above equation infinity, we obtain
\begin{eqnarray}
\lim_{R \to \infty}\lim_{W\to \infty}
\frac{1}{2R(2R+1)^{d-1}}\langle 
[-i \sum_{{\bf x}\in [-W,W]^d} x_1 n({\bf x}),H_R]\rangle
=\langle j({\bf 0},{\bf e}_1)\rangle.
\label{t=0}
\end{eqnarray}
The ordering of the above limiting procedures is 
crucial and can not be exchanged. In fact it is easy to see 
that if we take $R\to \infty$ first, for steady states 
it always gives zero. Moreover we can show the 
following lemma.
\begin{lemma}\label{lemma1}
For an arbitrary translationally invariant state and for any time 
$t\in {\bf R}$,
 \begin{eqnarray*}
\lim_{R \to \infty}\lim_{W\to \infty}
\frac{1}{2R(2R+1)^{d-1}}
\langle [-i \sum_{{\bf x}\in [-W.W]^d} x_1 n({\bf x}),\alpha_t(H_R)]\rangle
=\langle j({\bf 0},{\bf e}_1)\rangle
\end{eqnarray*}
holds, where $\alpha_t(H_R)$ is a 
time evolved object of $H_R$ (remind that 
we are in the Heisenberg picture).
Although the lhs includes time $t$ explicitly, its value does not
depend on $t$. 
\end{lemma}
{\bf Proof:}
\\
Since for $t=0$, the above equation holds (see (\ref{t=0})), 
we estimate its 
deviation for finite $t$.
That is, 
since 
\begin{eqnarray*}
&&
\frac{1}{2R(2R+1)^{d-1}}
[-i \sum_{{\bf x}\in [-W,W]^d} x_1 n({\bf x}),\alpha_t(H_R)] \nonumber \\
&=&\frac{1}{2R(2R+1)^{d-1}}[-i \sum_{{\bf x}\in [-W,W]^d} x_1 n({\bf x}),H_R]
\nonumber \\
&+&\frac{1}{2R(2R+1)^{d-1}}
\int^t_0 ds [-i \sum_{{\bf x}\in [-W,W]^d} x_1 n({\bf x}), 
-i\alpha_s([H_R,H_{R+1}])]
\nonumber
\end{eqnarray*}
holds, to show 
\begin{eqnarray}
\lim_{R \to \infty} \lim_{W \to \infty}
\frac{1}{2R(2R+1)^{d-1}}\Vert [-i \sum_{{\bf x}\in [-W,W]^d} x_1 n({\bf x}), 
-i\alpha_s([H_R,H_{R+1}])]\Vert 
=0
\nonumber \\
\label{s}
\end{eqnarray}
proves our lemma.
Let us consider first the case $s=0$ in the above equation.
\begin{eqnarray}
&&
\frac{1}{2R(2R+1)^{d-1}}[-i \sum_{{\bf x}\in [-W,W]^d} x_1 n({\bf x}), 
-i[H_R,H_{R+1}]]\nonumber \\
&=&\frac{1}{2R(2R+1)^{d-1}}
[H_R,[H_{R+1},\sum_{{\bf x}\in [-W,W]^d} x_1 n({\bf x})]]
\nonumber \\
&+&\frac{1}{2R(2R+1)^{d-1}}[H_{R+1},[\sum_{{\bf x}\in [-W,W]^d} x_1 n({\bf x}), H_R]]
\label{s=0}
\end{eqnarray}
which is obtained by Jacobi identity can be rewritten as, by letting 
$W \to \infty$,
\begin{eqnarray*}
\lim_{W\to \infty} (\ref{s=0})
&=& \frac{1}{2R(2R+1)^{d-1}}\left\{ [H_{R+1},-i \sum_{\{{\bf x},{\bf x}+{\bf e}_1\}
\subset [-R-1,R+1]^d \setminus [-R,R]^d} j({\bf x},{\bf x}+{\bf e}_1)]
\right.
\nonumber \\
&&
\left.
- [H_{R+1}-H_R, -i \sum_{\{{\bf x},{\bf x}+{\bf e}_1\}
\subset [-R-1,R+1]^d} j({\bf x},{\bf x}+{\bf e}_1)]\right\},
\end{eqnarray*}
whose norm is estimated as $O(1/R)$ and goes to zero as $R\to \infty$.
Next we consider the case of finite $s$ in (\ref{s}).
Thanks to 
\begin{eqnarray*}
&& \frac{1}{2R(2R+1)^{d-1}}
\Vert [-i \sum_{{\bf x}\in [-W,W]^d} x_1 n({\bf x}), 
-i\alpha_s([H_R,H_{R+1}])]\Vert
\nonumber \\
&=&
\frac{1}{2R(2R+1)^{d-1}}
\Vert [-i \sum_{{\bf x}\in [-W,W]^d} x_1 \alpha_{-s}(n({\bf x})), 
-i[H_R,H_{R+1}]]\Vert
\nonumber \\
&\leq &
\frac{1}{2R(2R+1)^{d-1}}
\Vert [-i \sum_{{\bf x}\in [-W,W]^d} x_1 n({\bf x}), 
-i[H_R,H_{R+1}]]\Vert
\nonumber \\
&+& \frac{1}{2R(2R+1)^{d-1}}
\left| 
\int^0_{-s} du 
\Vert [-i \sum_{{\bf x}\in [-W,W]^d} x_1 \frac{d \alpha_u(n({\bf x}))}{du}, 
-i[H_R,H_{R+1}]]\Vert
\right|,
\end{eqnarray*} 
to show that the term 
\begin{eqnarray}
\frac{1}{2R(2R+1)^{d-1}}
\Vert [-i \sum_{{\bf x}\in [-W,W]^d} x_1 \frac{d \alpha_u(n({\bf x}))}{du}, 
-i[H_R,H_{R+1}]]\Vert
\label{u}
\end{eqnarray}
for each $u$ 
converges to zero as $W \to \infty$ and $R \to \infty$ 
proves the claim.
Since $\frac{d \alpha_u(n({\bf x}))}{du}=\sum_{l=1}^d 
(\alpha_u(j({\bf x}-{\bf e}_l, {\bf x}))
-\alpha_u(j({\bf x},{\bf x}+{\bf e}_l)))
$ holds, it follows that 
\begin{eqnarray}
&&\sum_{{\bf x}\in [-W,W]^d} x_1 \frac{d \alpha_u(n({\bf x}))}{du}
\nonumber \\
&=&
\sum_{-W \leq x_2,\cdots,x_d \leq W}
\left\{
-W \alpha_u(j(-W-1,x_2,\cdots,x_d,-W,x_2,\cdots,x_d))\right.
\nonumber \\
&-&W \alpha_u(j(W,x_2,\cdots,x_d,W+1,x_2,\cdots,x_d))
+\sum_{x_1=-W}^{W-1}\left.
\alpha_u(j({\bf x},{\bf x}+{\bf e}_1))\right\}
\nonumber \\
&+&
\sum_{l=2}^d
\sum_{-W\leq x_1,x_2,\cdots,x_{l-1},x_{l+1},\cdots,x_d \leq W}
\nonumber \\
&&
x_1\left\{\alpha_u(j(x_1,x_2,\cdots,x_{l-1},-W-1,x_{l+1},\cdots,x_d,
x_1,\cdots,-W,\cdots,x_d))\right.
\nonumber \\
&-&
\left.\alpha_u(j(x_1,x_2,\cdots,x_{l-1},W,x_{l+1},\cdots,x_d,
x_1,\cdots,W+1,\cdots,x_d))\right\}
\label{v}
\end{eqnarray}
holds. 
We employ repeatedly the following group velocity lemma
(Theorem 6.2.11 of \cite{BR}).
There exists a positive constant $V$ such that 
for strictly local observables $A$ and  $B$ 
the following inequality holds:
\begin{eqnarray*}
\Vert [A,\alpha_t\circ \tau_{\bf x}(B)]\Vert
\leq 
C \Vert A\Vert \Vert B \Vert
\exp \left(
-|x|+V|t|\right),
\end{eqnarray*}
where $C$ is a constant depending only on 
size of the regions where $A$ and $B$ live in. 
Let us substitute (\ref{v}) for (\ref{u})
and estimate its each term.
Thanks to $\Vert -i[H_R,H_{R+1}]\Vert =O(R^{d-1})$,
\begin{eqnarray*}
&&\frac{1}{2R(2R+1)^{d-1}}
\Vert [\sum_{-W \leq x_2,\cdots,x_d \leq W}
\nonumber \\
&&
-W \alpha_u(j(-W-1,x_2,\cdots,x_d,-W,x_2,\cdots,x_d))
, -i[H_R,H_{R+1}]]\Vert
\nonumber\\
&\leq&
O\left(\frac{1}{R}\right)
O(W^d) \exp\left(-(W-R) +V|u| \right)
\end{eqnarray*}
holds 
and it goes to zero as $W\to \infty$.
The commutator related with 
$-W \alpha_u(j(W,x_2,\cdots,x_d,W+1,x_2,\cdots,x_d))$
also vanishes in the same manner.
Next one can see that 
\begin{eqnarray*}
&&\frac{1}{2R(2R+1)^{d-1}}
\Vert
[\sum_{x_1=-W}^{W-1}\sum_{-W \leq x_2,\cdots,x_d \leq W}
\alpha_u(j({\bf x},{\bf x}+{\bf e}_1)),
-i[H_R,H_{R+1}]]\Vert
\nonumber \\
&\leq&
\left( \sum_{{\bf x} \in {\bf Z}^d}
e^{-|{\bf x}|}\right)e^{V|u|}
O\left( \frac{1}{R}\right)
\end{eqnarray*}
holds and it goes to zero as $R\to \infty$.
Finally,
\begin{eqnarray*}
&&\frac{1}{2R(2R+1)^{d-1}}
\Vert [
\sum_{-W\leq x_1,x_2,\cdots,x_{l-1},x_{l+1},\cdots,x_d \leq W}
\nonumber \\
&&
x_1 \alpha_u(j(x_1,x_2,\cdots,x_{l-1},-W-1,x_{l+1},\cdots,x_d,
x_1,\cdots,-W,\cdots,x_d)), 
\nonumber \\
&&
-i[H_R,H_{R+1}]]\Vert
\nonumber \\
&\leq&
O\left(\frac{1}{R}\right)
O(W^{d})
e^{-(W-R)+V|u|}
\end{eqnarray*}
holds and it goes to zero as $W \to \infty$ and the 
term related with $x_l=W$ vanishes similarly.
Thus we proved the equation,
 \begin{eqnarray*}
\lim_{R \to \infty}\lim_{W\to \infty}
\frac{1}{2R(2R+1)^{d-1}}\langle 
[-i \sum_{{\bf x}\in [-W,W]^d} x_1 n({\bf x}),H_R(t)]\rangle
=\langle j({\bf 0},{\bf e}_1)\rangle.
\end{eqnarray*}
\hfill Q.E.D.
\par
By using an expression $H_R=\sum_{r=1}^d
\sum_{-R\leq y_r \leq R-1}\sum_{-R \leq y_1,\cdots,y_{r-1},y_{r+1},
\cdots,y_d \leq R}h({\bf y},{\bf y}+{\bf e}_r)$, one obtains 
for translationally invariant states,
\begin{eqnarray*}
\langle j({\bf 0},{\bf e}_1)\rangle
&=&\lim_{R\to \infty}
\frac{-i}{2R(2R+1)^{d-1}}
\sum_{{\bf x} \in {\bf Z}^d}
\sum_{r=1}^d
\sum_{-R\leq y_r \leq R-1}\sum_{-R \leq y_1,\cdots,y_{r-1},y_{r+1},
\cdots,y_d \leq R}
\nonumber \\
&&
x_1 \langle [n({\bf x}),\alpha_t(h({\bf y},{\bf y}+{\bf e}_r))]\rangle
\end{eqnarray*}
from lemma \ref{lemma1}.
Thanks to the translational invariance of the state,
substitution ${\bf z}:={\bf y}-{\bf x}$ makes the above equation
\begin{eqnarray*}
\langle j({\bf 0},{\bf e}_1)\rangle
&=&
\lim_{R\to \infty}
\frac{-i}{2R(2R+1)^{d-1}}
\sum_{{\bf z} \in {\bf Z}^d}
\sum_{r=1}^d
\sum_{-R\leq y_r \leq R-1}\sum_{-R \leq y_1,\cdots,y_{r-1},y_{r+1},
\cdots,y_d \leq R}
\nonumber \\
&&
(y_1 -z_1) \langle [n({\bf 0}),
\alpha_t(h({\bf z},{\bf z}+{\bf e}_r))]\rangle.
\end{eqnarray*}
One can perform summation for ${\bf y}$ 
of the above equation.
For $r=1$, one obtains
\begin{eqnarray*}
\frac{-i}{2R(2R+1)^{d-1}}
\sum_{-R\leq y_1 \leq R-1}\sum_{-R \leq y_2,\cdots,y_d \leq R}
(y_1 -z_1) 
=i\left( z_1 +\frac{1}{2} \right),
\end{eqnarray*}
and
for $r\neq 1$, 
\begin{eqnarray*}
\frac{-i}{2R(2R+1)^{d-1}}
\sum_{-R\leq y_r \leq R-1}\sum_{-R \leq y_1,\cdots,y_{r-1},y_{r+1},
\cdots,y_d \leq R}
(y_1 -z_1) 
=iz_1.
\end{eqnarray*}
Thus we obtain
\begin{eqnarray}
\langle j({\bf 0},{\bf e}_1)\rangle
&=&
i \sum_{{\bf z}\in {\bf Z}^d}
\left\{
\left( z_1 +\frac{1}{2} \right) 
\langle [n({\bf 0}), \alpha_t(h({\bf z},{\bf z}+{\bf e}_1))]\rangle
\right.
\nonumber \\
&&
+
\left. i\sum_{r=2}^d z_1 
\langle [n({\bf 0}), \alpha_t(h({\bf z},{\bf z}+{\bf e}_r))]\rangle\right\}.
\label{rev}
\end{eqnarray}
Thanks to this representation, 
we obtain the following theorem.
\begin{theorem}
For translationally invariant 
states with nonvanishing current, there exists at least 
one $r \in \{1,2,\cdots,d\}$
such that  
neither $\langle [n({\bf 0}),\alpha_t(h({\bf z},{\bf z}+{\bf e}_r))]
\rangle$ nor a dynamical correlation function 
\begin{eqnarray*}
\langle n({\bf 0})\alpha_t(h({\bf z},{\bf z}+{\bf e}_r))\rangle^{T}
:=\langle n({\bf 0})\alpha_t(h({\bf z},{\bf z}+{\bf e}_r))\rangle
-\langle n({\bf 0})\rangle
\langle \alpha_t(h({\bf z},{\bf z}+{\bf e}_r))\rangle
\end{eqnarray*}
are absolutely integrable with respect to 
${\bf z}$ and $t$.
\end{theorem}
{\bf Proof:}
First let us note that if $\langle 
[n({\bf 0}),\alpha_t(h({\bf z},{\bf z}+{\bf e}_r))]
\rangle$
is not absolutely integrable, 
$\langle n({\bf 0})\alpha_t(h({\bf z},{\bf z}+{\bf e}_r))\rangle^{T}$
is not also absolutely integrable.
In fact thanks to 
\begin{eqnarray*}
\langle n({\bf 0})\alpha_t(h({\bf z},{\bf z}+{\bf e}_r))\rangle^{T}
-
\overline{
\langle n({\bf 0})\alpha_t(h({\bf z},{\bf z}+{\bf e}_r))\rangle^{T}}
=\langle 
[n({\bf 0}),\alpha_t(h({\bf z},{\bf z}+{\bf e}_r))]
\rangle,
\nonumber
\end{eqnarray*}
$|\langle 
[n({\bf 0}),\alpha_t(h({\bf z},{\bf z}+{\bf e}_r))]
\rangle|
\leq 2 
|\langle n({\bf 0})\alpha_t(h({\bf z},{\bf z}+{\bf e}_r))\rangle^{T}|$
is obtained.
Now we show that there exists $r$ such that $\langle 
[n({\bf 0}),\alpha_t(h({\bf z},{\bf z}+{\bf e}_r))]
\rangle$ is not absolutely integrable.
From (\ref{rev}), one can see that
\begin{eqnarray*}
|\langle j({\bf 0},{\bf e}_1)\rangle|
&\leq &
\sum_{{\bf z}\in {\bf Z}^d}\left\{
\left| z_1 +\frac{1}{2} \right|
| \langle [n({\bf 0}), \alpha_t(h({\bf z},{\bf z}+{\bf e}_1))]\rangle |
\right.
\nonumber \\
&&
+
\left.
\sum_{r=2}^d | z_1 |
| 
\langle [n({\bf 0}), \alpha_t(h({\bf z},{\bf z}+{\bf e}_r))]\rangle |
\right\}
\nonumber \\
&\leq&
\sum_{{\bf z}\in {\bf Z}^d}
\left( | z_1 | +\frac{1}{2} \right)
\sum_{r=1}^d
| \langle [n({\bf 0}), \alpha_t(h({\bf z},{\bf z}+{\bf e}_r))]\rangle |
\end{eqnarray*}
holds.
Thanks to the group velocity lemma,
there exists a constant $C$ and $V$ such that
\begin{eqnarray*}
|\langle 
[n({\bf 0}),\alpha_t(h({\bf z},{\bf z}+{\bf e}_r))]
\rangle|
\leq C \Vert n({\bf 0})\Vert 
\Vert h({\bf 0},{\bf 0}+{\bf e}_r)\Vert
\exp \left( -|{\bf z}|+V|t|\right)
\nonumber
\end{eqnarray*}
holds.
Therefore, for an arbitrary $\epsilon >0$,
there exists $N>0$ satisfying 
\begin{eqnarray*}
\sum_{|{\bf z}|>V|t|+N} \left( | z_1 | +\frac{1}{2}\right)
\sum_{r=1}^d
|\langle 
[n({\bf 0}),\alpha_t(h({\bf z},{\bf z}+{\bf e}_r))]
\rangle| < \epsilon.
\end{eqnarray*}
Thus we obtain
\begin{eqnarray*}
|\langle j({\bf 0},{\bf e}_1)\rangle|
&\leq&
\sum_{{\bf z}\in {\bf Z}^d}
\left( | z_1 | +\frac{1}{2} \right)
\sum_{r=1}^d
| \langle [n({\bf 0}), \alpha_t(h({\bf z},{\bf z}+{\bf e}_r))]\rangle |
\nonumber \\
&\leq &
\epsilon+\left\{ V|t|+N+\frac{1}{2}\right\}\sum_{r=1}^d 
\sum_{|{\bf z}|\leq V|t|+N}
|\langle [n({\bf 0}),\alpha_t(h({\bf z},{\bf z}+{\bf e}_r))]
\rangle|
\nonumber \\
&\leq&
\epsilon +\left( V|t|+N+\frac{1}{2}\right)
\sum_{r=1}^d
\sum_{{\bf z}\in {\bf Z}^d}
|\langle [n({\bf 0}),\alpha_t(h({\bf z},{\bf z}+{\bf e}_r))]
\rangle|,
\end{eqnarray*}
which leads to 
\begin{eqnarray*}
\sum_{r=1}^d
\sum_{{\bf z}\in {\bf Z}^d}
|\langle [n({\bf 0}),\alpha_t(h({\bf z},{\bf z}+{\bf e}_r))]\rangle|
\geq \frac{|\langle j({\bf 0},{\bf e}_1)\rangle|-\epsilon}
{V|t|+N+1/2}.
\end{eqnarray*}
If for all $r=1,2,\cdots,d$, 
$|\langle [n({\bf 0}),\alpha_t(h({\bf z},{\bf z}+{\bf e}_r))]\rangle|$
are integrable,
their summation 
$\sum_{r=1}^d 
|\langle [n({\bf 0}),\alpha_t(h({\bf z},{\bf z}+{\bf e}_r))]\rangle|$
must be integrable. 
It ends the proof.
\hfill Q.E.D.
\section{Conclusion and Outlook}
In this paper, 
we considered states over $d$-dimensional infinite spin systems
which are translationally invariant and have non-vanishing
current expectations. 
The dynamical correlation function between $n({\bf 0})$ and 
$\alpha_t(h({\bf x},{\bf x}+{\bf e}_r))$ for some $r$ shows 
slow decay with respect to space and time $({\bf x},t)$.
The key observation to prove our theorem is 
to express the current operator by a commutator
and to apply the argument of $\cite{Req}$ for the 
continuous symmetry breaking case to our nonequilibrium case,
although no continuous symmetry breaking takes place 
in our case. 
The behavior of 
dynamical correlation function 
is related with observable effect through 
response functions, which will be 
discussed elsewhere. 
Since the conditions for our nonequilibrium states 
are weak, our result is 
general but is not very strong. 
It is interesting to investigate whether 
additional physical conditions can derive more 
detailed form of the correlation function.
\\
{\bf Acknowledgment}
\\
The author would like to thank anonymous referees for 
fruitful comments.
\\


\begin{thebibliography}{99}
\bibitem{energycurrent}
In the absence of charge operator,
energy current can be defined from local Hamiltonian.
The discussion in this paper holds also for the case
when states have nonvanishing energy current.
\bibitem{real}
In real experimental circumstance, nonequilibrium states 
are often realized by a several reservoirs attached to 
boundary of a finite system. The systems considered in 
this paper do not have boundary and might be 
regarded as a ``thermodynamic limit" of such a 
real system. We do not know conditions which 
justify this point of view.
\end{thebibliography}
\end{document}